# Prolonging Carrier Lifetime in P-type 4H-SiC Epilayer by Thermal Oxidation and Hydrogen Annealing

**Ruijun Zhang[1], Mingkun Zhang[3], Guoliang zhang[1], Yujian Chen[1], Jia Liu[1], Ziqian Tian[1], Ye Yu[1*], Peng zhao[1], Shaoxiong Wu[1], Yuning Zhang[1], Dingqu Lin[1], Xiaping Chen[1], Jiafa Cai[1], Rongdun Hong[1,2 *], Feng Zhang [1,2*]**

[1]Department of physics, Xiamen university, Fujian, 361005, P. R. China
[2]Jiujiang Research Institute of Xiamen University, Jiangxi 332000, P. R. China
[3]The Higher Educational Key Laboratory of Flexible Manufacturing Equipment Integration of Fujian Province, Xiamen Institute of Technology, Fujian, 361005, P. R. China
E-mail: fzhang@xmu.edu.cn, rdhong@xmu.edu.cn, yu.ye@sicty.com

**Abstract**

A minority carrier lifetime of 25.46 μs in a P-type 4H-SiC epilayer has been attained through sequential thermal oxidation and hydrogen annealing. Thermal oxidation can enhance the minority carrier lifetime in the 4H-SiC epilayer by reducing carbon vacancies. However, this process also generates carbon clusters with limited diffusivity and contributes to the enlargement of surface pits on the 4H-SiC. High-temperature hydrogen annealing effectively reduces stacking fault and dislocation density. Moreover, electron spin resonance analysis indicates a significant reduction in carbon vacancy defects after hydrogen annealing. The mechanisms of the elimination of carbon vacancies by hydrogen annealing include the decomposition of carbon clusters formed during thermal oxidation and the low-pressure selective etching by hydrogen, which increases the carbon content on the 4H-SiC surface and facilitates carbon diffusion. Consequently, the combination of thermal oxidation and hydrogen annealing eliminates carbon vacancies more effectively, substantially enhancing the minority carrier lifetime in P-type 4H-SiC. This improvement is advantageous for the application of high-voltage SiC bipolar devices.

## Introduction

Silicon carbide (SiC) has received increasing attention and been widely applied in high-temperature, high-power, and high-frequency electronic devices owing to its superior physical properties such as wide band gap (3.26 eV), high carrier saturation rate (1200 $cm^2v^{-1}s^{-1}$) and good thermal conductivity (4.9 $Wcm^{-1}K^{-1}$)[1, 2]. For ultrahigh-voltage applications (>10 kV), bipolar devices have more absolute advantages than unipolar devices in terms of lower on-resistance owing to the effect of conductivity modulation[3, 4]. The ultrahigh-voltage bipolar devices require a long carrier lifetime to achieve sufficient conductivity modulation for high current density [5]. At present, one of the dominant obstacles to realizing high-current SiC bipolar devices is the low minority carrier lifetimes in SiC epilayers[3, 6]. The typical minority carrier lifetime observed in 4H-SiC epilayer as deposited is usually less than 1 μs , while 5 μs is required for 10 kV bipolar devices[7]. Therefore, in order to improve the minority carrier lifetime, it is an important development orientation to identify and reduce corresponding killer defects in the SiC epilayer [8, 9].

The deep-level defects in the 4H-SiC epilayer, usually working as a recombination center, have been intensively investigated by several groups[10-12]. The $Z_{1/2}$ center is currently recognized as the primary lifetime-killing defect in lightly doped N-type 4H-SiC epilayer[13, 14], which is located at 0.6-0.7 eV below the conduction band edge and considered to be the recombination center related to carbon vacancies ($V_c$). The growth processing of SiC epilayers had quite a significant impact on the density of the $Z_{1/2}$ center to regulate carrier lifetime[15]. Therefore, it is meaningful to decrease the density of the $Z_{1/2}$ during epitaxy growth by optimizing the C/Si ratio and temperature[16]. Moreover, thermal oxidation or carbon ion implantation during post-growth can eliminate $Z_{1/2}$ defect centers to obtain long minority carrier lifetimes of 20-30 μs in N-type 4H-SiC epilayers with a thickness of 200-350um[17-19]. However, it has been found that the application of thermal oxidation or carbon ion implantation is not effective adequately in improving the minority carrier lifetime of P-type 4H-SiC epilayers[20, 21]. Although extensive research has been carried out on N-type 4H-SiC, there is few specific study to elucidate the dominant mechanism of carrier recombination in P-type epilayer[22]. $EH_{6/7}$ center rather than the $Z_{1/2}$ center must be more involved in carrier recombination in P-type SiC [23]. Takafumi Okuda et al reported an improvement of minority carrier lifetimes in P-type 4H-SiC epilayer by hydrogen annealing[24]. But it is still unknown that the mechanism of hydrogen annealing on improving minority carrier lifetimes.

In this study, the minority carrier lifetime of P-type 4H-SiC epilayer before and after thermal oxidation and hydrogen annealing were studied by using microwave photoconductive decay (μ-PCD). The adverse effect of thermal oxidation on the minority carrier lifetime of specific P-type 4H-SiC epilayer was illustrated by analyzing the change of defects before and after oxidation. Additionally, deep level transient spectroscopy (DLTS) and electron spin resonance (ESR) were used to characterize and analyze the deep level defects. This paper discusses in detail the physical mechanisms underlying the enhancement of minority carrier lifetime in 4H-SiC by high-temperature oxidation and hydrogen annealing

**Experiments**

The samples used in this study were P-type 4° off-axis 4H-SiC (0001) epilayer grown on N-type substrates. The thickness and doping concentration of the epilayer were 150 μm and $2 \times 10^{14}$ cm$^{-3}$, respectively. After standard RCA cleaning, thermal oxidation at 1300°C for 8 hours was conducted on epitaxial layers with different defect densities. Minority carrier lifetime and photoluminescence (PL) defect assessments were performed before and after the oxidation. In order to investigate the effect of hydrogen annealing process on stacking faults and dislocations in 4H-SiC, high-temperature hydrogen annealing was conducted at 1000 °C to 1650 °C for 0.5, 1, and 1.5 hours.

Combining the advantages of thermal oxidation and hydrogen annealing, a dual process was developed to enhance the minority carrier lifetime in P-type 4H-SiC. The samples were thermally oxidized at 1300 ℃ for 8 h (1st thermal oxidation) and applied on annealing in hydrogen at 1000 °C for 1 hour under low pressure (1st hydrogen annealing), by using the low-pressure chemical vapour deposition (LPCVD) equipment. After 1 hour hydrogen annealing, the samples were thermally oxidized again at 1300 °C for 16 h (2nd thermal oxidation), followed by hydrogen annealing again at 1000 °C for 2 h (2nd hydrogen annealing). All hydrogen annealing processes were conducted under sub-atmospheric pressures. The dislocation defects of the epitaxial wafers were observed using photoluminescence (PL) and bright filed (BF) before and after the treatment.

The μ-PCD was measured at room temperature by Semilab WT-2000 instrument to obtain the curves of photoconductivity decay and the whole epilayer mappings of minority carrier lifetimes. During the measurement, the samples were excited by an Yttrium Lithium Fluoride (YLF) laser with a wavelength of 349 nm and a pulse of 15 ns. The decay of photoconductivity was monitored via microwave reflectivity at 26 GHz to assess variations in excess carrier concentration. In this study, the minority carrier lifetime decay curves and lifetime mapping are represented by the measured lifetime $\tau_{eff}$, from which the bulk lifetime $\tau_{bulk}$ can be obtained through exponential fitting. To investigate the variations in defect distributions induced by hydrogen annealing, a P-type epitaxial wafer was annealed in a hydrogen gas environment at 1000°C for 3 hours. It was then analyzed using DLTS and ESR both before and after the annealing process. For ESR test, spectra were recorded at a resonant frequency of 9.3626 GHz with a Bruker EMX Plus X-band Continuous-Wave (CW) EPR spectrometer. The spectrometer was set to the following parameters for all experiments: microwave power of 20 mW, modulation amplitude of 1 G, modulation frequency of 100 kHz, sweep width of 300 G, and temperature of 123 K. For DLTS testing, the samples were prepared as follows. Firstly, an Ohmic contact layer was established on the surface of p-type epitaxial layer by surface-selective Al$^+$ ion implantation. Subsequently, activation annealing was performed at 1650°C for 30 minutes in an Ar atmosphere. After surface oxide etching using BOE, Ti/Al/Ti/Au layers were magnetron sputtered on the implantation area, followed by rapid thermal annealing (RTA) at 1000°C for 1 minute to form robust Ohmic contacts. Concurrently, Ti/Au layers were deposited on the non-implanted area to form Schottky contacts.

**Results and discussion**

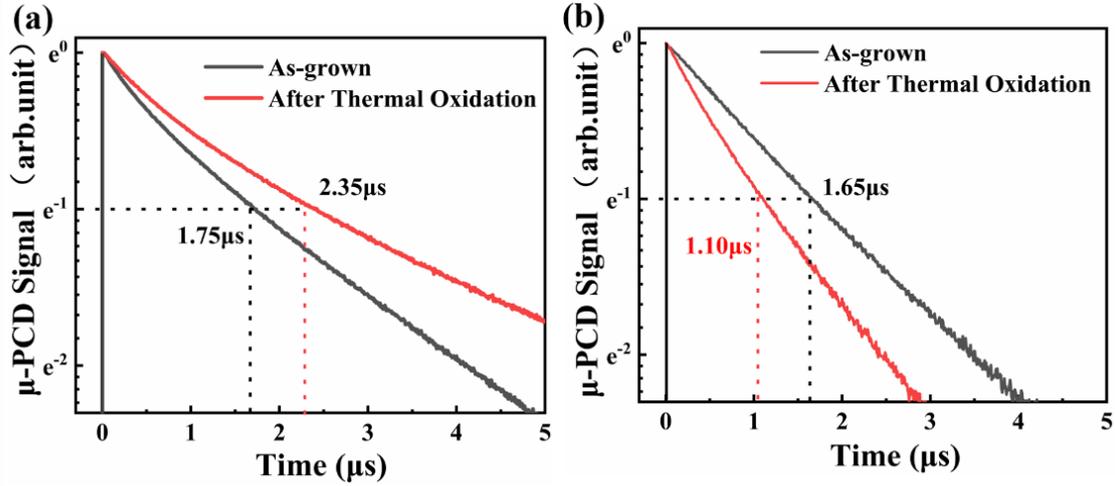

Figure. 1 The μ-PCD decay curves at room temperature for P-type 4H-SiC epilayer with different defect density before (as-grown) and after thermal oxidation. (a) Surface defects density:1020 cm$^{-2}$; (b) Surface defect:331 cm$^{-2}$

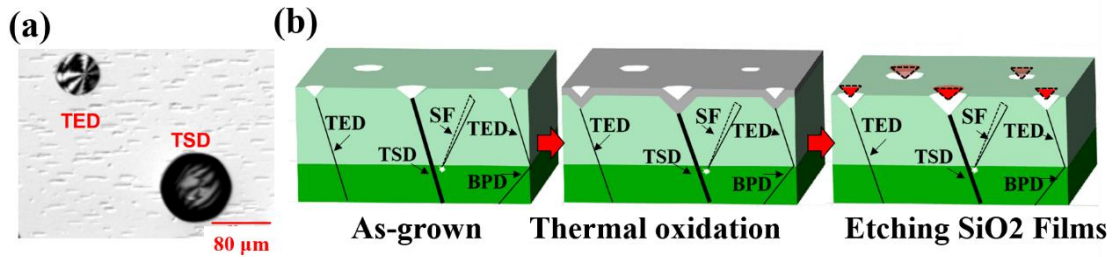

Figure. 2 (a) Morphology of defects after KOH etching of P-type 4H-SiC epilayer. (b) Schematic diagrams of defects change before and after thermal oxidation.

## I. Thermal Oxidation

As shown in Figure 1(a), for the P-type 4H-SiC epilayer with low surface defects density of 331 cm$^{-2}$, the minority carrier lifetime increased from 1.75 μs to 2.35 μs under the same oxidation conditions. In previous work, as shown in Figure 1(b), the minority carrier lifetimes of P-type 4H-SiC epilayer with high surface defect density of 1020 cm$^{-2}$ decreased from 1.65 μs to 1.10 μs after thermal oxidation at 1300°C for 8 hours[25]. Similar reduction of minority carrier lifetimes after thermal oxidation has also been reported.[26] Therefore, a high density of surface defects reduces minority carrier lifetimes in SiC after thermal oxidation. Conversely, a low defect density increases minority carrier lifetimes. In prior studies, the evolution of photoluminescence (PL) white spot defects in P-type 4H-SiC epitaxial layers before and after thermal oxidation was systematically presented by PL) spectra images.[25] As shown in Figure 2(a), These white spot defects were further identified as threading screw dislocations (TSD) and threading edge dislocations (TED) through KOH etching experiments. Although many basal plane dislocations (BPDs) are converted to TEDs during the epitaxial growth process, BPDs can still be observed. [27, 28].

These dislocations form pits easily at their surface outcrops, exhibiting higher oxidation rates during thermal oxidation. After removing the oxide layer, these pits develop into larger surface etch pits, as illustrated in Figure 2(b). When the crystal quality is poor and the surface defects are numerous, the improvement of minority carrier lifetime caused by thermal oxidation will be restrained. Consequently, higher quality epilayers with fewer surface defects are required in order to improve the minority carrier lifetime by thermal oxidation. Additionally, no significant changes in the morphology or quantity of stacking faults (SF) and BPD were observed in the PL channel after thermal oxidation.

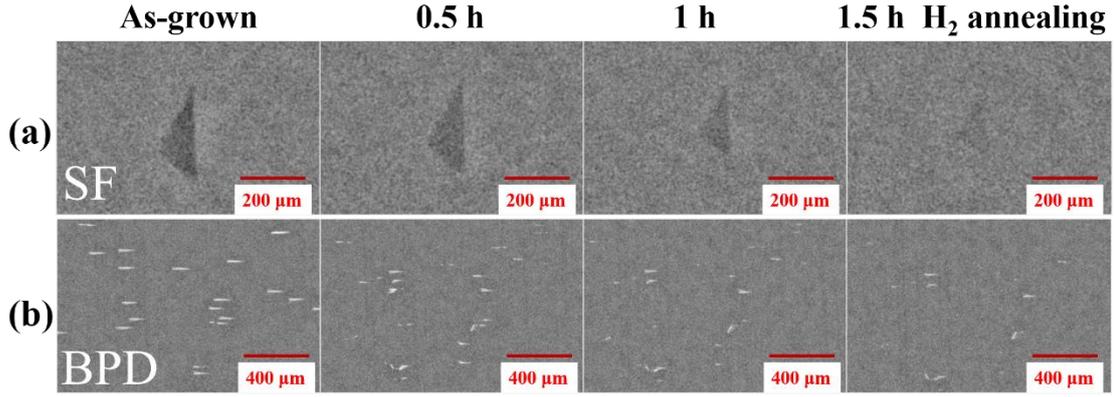

Figure. 3 PL images of (a) SF and (b) BPD defects before and after hydrogen annealing.

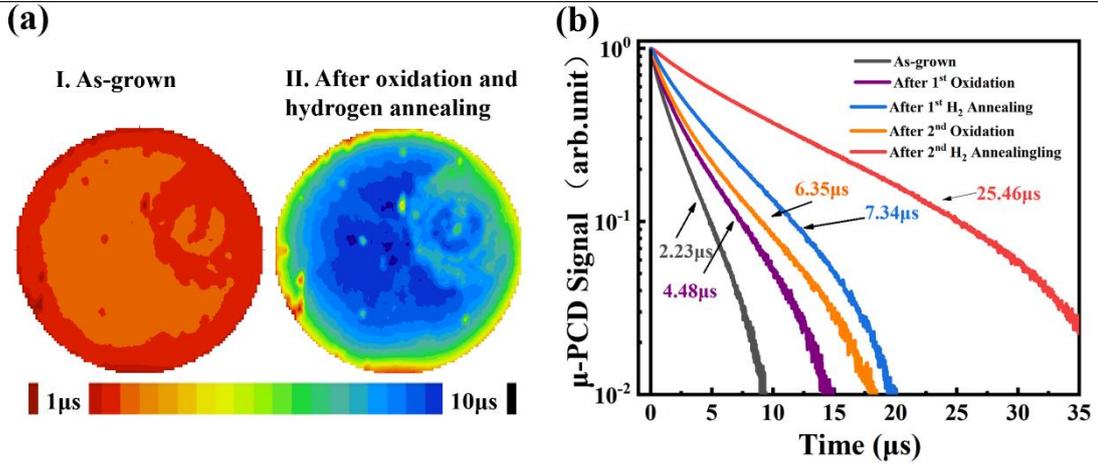

Figure.4(a) The μ-PCD curves measured at room temperature for P-type 4H-SiC epilayer before (as-grown) and after double thermal oxidation and hydrogen annealing. (b) Minority carrier lifetimes mapping measured by μ-PCD for the 4-inch P-type 4H-SiC wafer before (as-grown) and after oxidation and hydrogen annealing twice.

## *II*. Hydrogen Annealing

High-temperature hydrogen annealing is performed to reduce stacking faults and dislocation density in 4H-SiC epilayers. As shown in Figure 3(a) and (b), changes in the morphology and quantity of SF and BPD on the epitaxial wafer are observed in the PL channel before and after hydrogen annealing. SFs are gradually eliminated after 1.5 hours of hydrogen annealing, while most BPDs are removed after 1 hour, with a few BPDs persisting after 1.5 hours. The etching effect of hydrogen annealing at high temperatures significantly enhances on silicon carbide. The etching of the surface layer may remove dislocations and stacking faults. However, the motion of basal plane dislocations (BPDs) can also be clearly observed after high-temperature annealing. Therefore, dislocations may recombine due to their motion at elevated temperatures. Stacking faults may be rearranged through atomic diffusion and reconstruction. These observations suggest that hydrogen annealing enhances the quality of the crystal. However, this phenomenon occurs only at elevated hydrogen annealing temperatures between 1200°C and 1650°C.

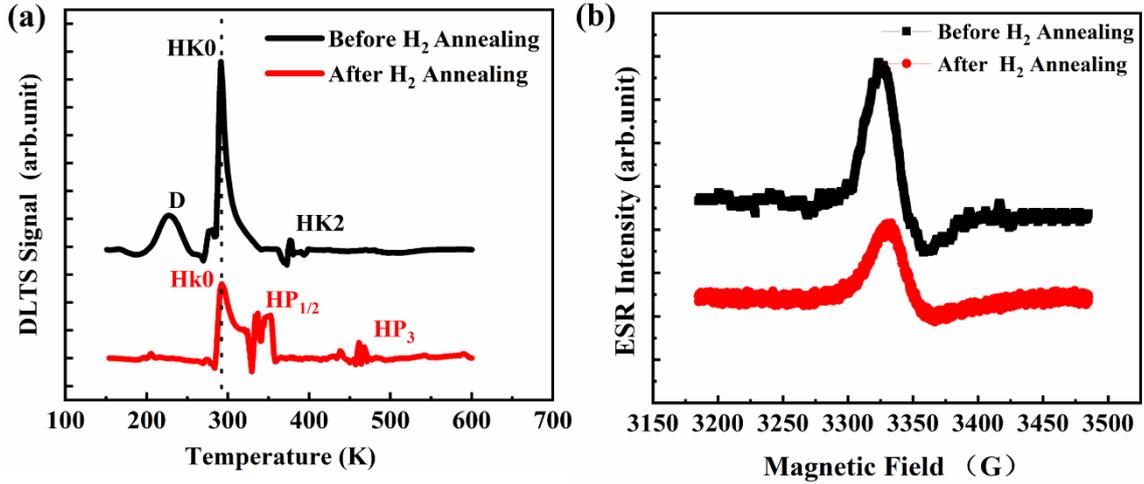

Fig. 5 (a) DLTS spectra of p-type 4H-SiC epilayer before and after hydrogen annealing. (b) ESR spectra of p-type 4H-SiC epilayer before and after hydrogen annealing at 123 K.

## III. Thermal Oxidation and Hydrogen Annealing

Building on the improvement of minority carrier lifetime in low-defect-density epitaxial wafers via thermal oxidation, hydrogen annealing was subsequently performed to further enhance the minority carrier lifetime. The mapping of carrier lifetime, measured by μ-PCD, reached 10μs after a combination of thermal oxidation and long-term hydrogen annealing, as shown in Figure 4(a). As shown in Equation (1), which describes the relationship among the effective minority carrier lifetime ($\tau_{eff}$), the bulk lifetime ($\tau_{bulk}$), suface recombination lifetime($\tau_{suf}$), and substrate recombination lifetime($\tau_{sub}$), $\tau_{eff}$ can be expressed as a sum of these contributions. The decay curve of the minority carrier lifetime can be modeled using a single-exponential fit to represent the combined effects of various recombination mechanisms. As shown in Equation (2), I(t) denotes the measured μ-PCD signal, and τ represents the minority carrier lifetime. During the fitting process, the lifetime τ corresponds to $\tau_{eff}$.

$$\frac{1}{\tau_{eff}} = \frac{1}{\tau_{suf}} + \frac{1}{\tau_{bulk}} + \frac{1}{\tau_{sub}} \quad (1)$$

$$I(t) = A\exp\left(\frac{-t}{\tau}\right) \quad (2)$$

$$I(t) = A_1 \exp\left(\frac{-t}{\tau_1}\right) + A_2 \exp\left(\frac{-t}{\tau_2}\right) + A_3 \exp\left(\frac{-t}{\tau_3}\right) \quad (3)$$

As shown in Equation (3), the bulk lifetime is more accurately determined using a triple-exponential fitting to the minority carrier lifetime decay curve. In device applications, bulk lifetime emerges as the paramount factor. This method enables the extraction of three distinct lifetimes, namely $\tau_1$, $\tau_2$, and $\tau_3$, each associated with a specific recombination mechanism. As illustrated in Figure 4(b), the decay curve consists of three distinct phases. The first and third phases, represented by $\tau_1$ and $\tau_3$, correspond to fast recombination. The second phase, represented by $\tau_2$, corresponds to slower recombination. Surface recombination occurs more rapidly, as carriers recombine at the material surface. Therefore, $\tau_1$ represents the surface recombination lifetime, $\tau_{suf}$. Higher doping concentrations in the substrate make it more prone to Auger recombination. Thus, τ3 represents the Auger recombination lifetime, $\tau_{sub}$. In contrast, the epitaxial layer with lower doping concentrations is dominated by SRH (Shockley-Read-Hall) recombination, with τ2 corresponding to the bulk recombination lifetime $\tau_{bulk}$.[29, 30].

Figure 4(b) shows the μ-PCD curves of P-type 4H-SiC epilayer before and after double thermal oxidation and hydrogen annealing. Initially, the minority carrier lifetime increased from 2.23 μs to 4.48 μs after 1st thermal oxidation (@1300 °C 8h). This was followed by an increase to 7.34 μs after 1st hydrogen annealing (@1000 °C 1h). Subsequently, the durations of thermal oxidation and hydrogen annealing durations were extended to 16 and 2 hours, respectively. Following the 2nd thermal oxidation (@1300 °C 16h), the minority carrier lifetime decreased from 7.34 μs to 6.35 μs, due to the instability of hydrogen passivation from the 1st hydrogen annealing. Finally, after the 2nd hydrogen annealing (@1000 °C 2h), the minority carrier lifetime increased significantly from 6.35 μs to 25.46 μs. Therefore, it is obvious that the minority carrier lifetimes of P-type SiC was significantly improved by the secondary oxidation and hydrogen annealing process.

The analysis results of DLTS and ESR tests were then used to study the physical mechanisms of the increase in the minority carrier lifetimes for p type SiC epitaxy. As illustrated in Figure 5(a), DLTS measurements were conducted on the P-type 4H-SiC epilayer before and after 3-hour hydrogen annealing process. The D-defect center is a Boron-related defect, which is believed to possibly originate from the graphite components used for epitaxial growth[31]. The HK0 and HK2 defects, which emerged after thermal oxidation, have also been documented by Katsunori Danno[32]. HK0 is considered as a complex comprising interstitial carbon atoms formed after thermal oxidation. Following hydrogen annealing, the D center is no longer detectable, and the concentration of HK0 substantially diminishes. This indicates that hydrogen annealing significantly reduces the formation of carbon complexes following thermal oxidation. Furthermore, new defects labelled HP$_{1/2}$ and HP$_3$ emerge after hydrogen annealing. The implications of the new defects are still being explored. As illustrated in Figure 5(b), ESR measurements were conducted on the P-type 4H-SiC epilayer before and after 3-hour hydrogen annealing process. The g-factor of the sample spectral line, which reflects the type of defect, can be determined using Equation (4).

$$g = \frac{h\nu}{\beta H} \quad (4)$$

$h$ represents Planck's constant, $\beta$ denotes the Bohr magneton, $\nu$ is the resonance frequency, and H indicates the central magnetic field. The g factor of carbon vacancy in the 4H-SiC epilayer ranges between 1.9963 and 2.0081[33, 34]. The g factor of the as-grown P-type 4H-SiC epilayer is 1.9968, which turned to 1.9965 after hydrogen annealing for 3 hours. This indicates that carbon vacancy defects are the primary point defects both before and after hydrogen annealing. Following hydrogen annealing, the significant decrease in spectral intensity suggests a reduction in the concentration of carbon vacancy defects. This reduction in carbon vacancy defects may be the primary reason for the improvement in minority carrier lifetime through hydrogen annealing.

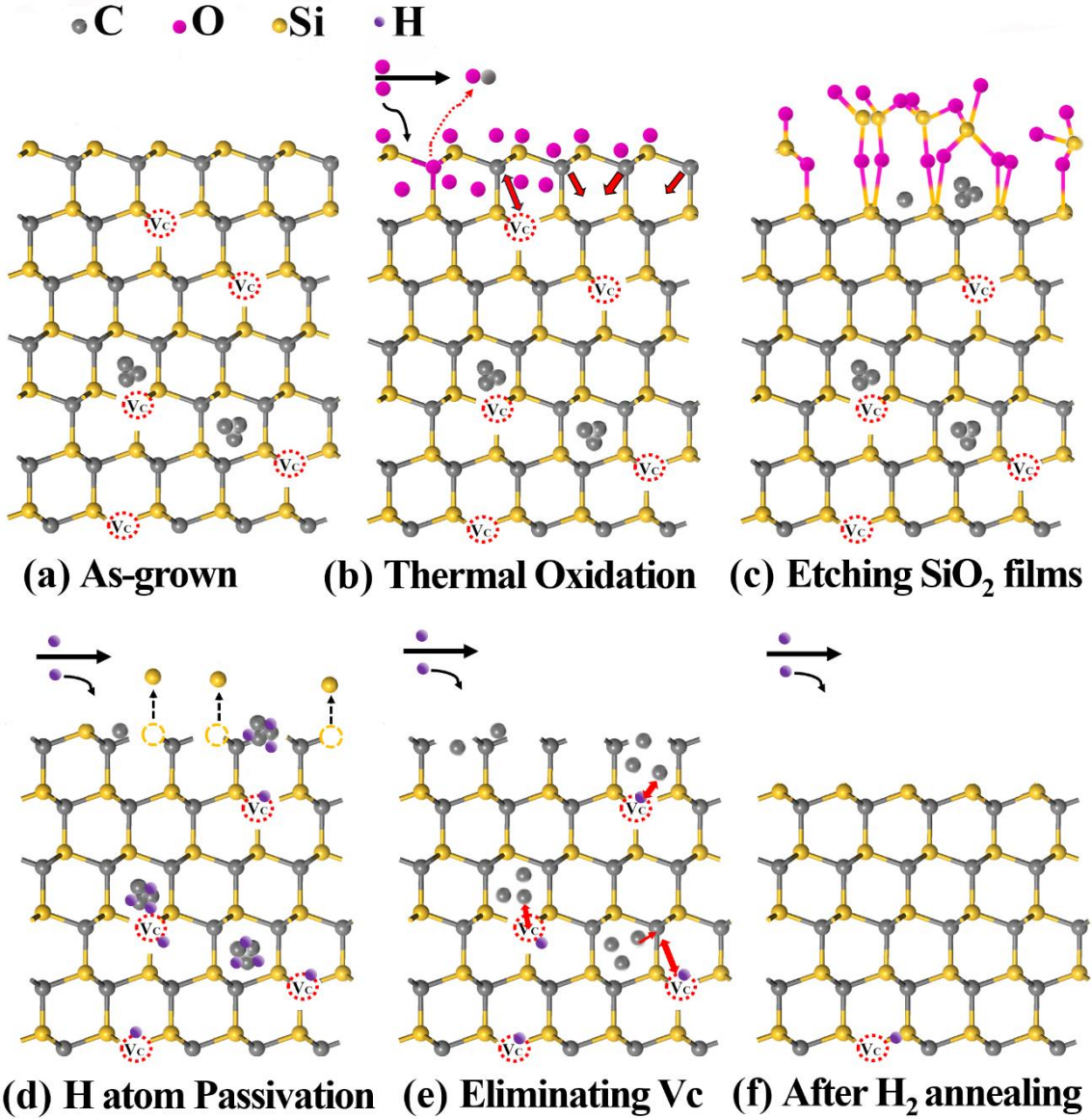

Fig.6 Schematic diagrams of the mechanism of thermal oxidation and hydrogen annealing to increase minority carrier lifetime. (a)as-grown and (b) thermal oxidation, (c) BOE etching oxide layer, (d-e) Atomic interactions during H$_2$ Annealing and (f) after H$_2$ annealing.

The physical mechanisms by which thermal oxidation and hydrogen annealing enhance minority carrier lifetime are discussed based on the measurement results from DLTS and ESR. As illustrated in Figure 6(a)-(c), Si atoms are oxidized to $SiO_2$ during thermal oxidation, while C atoms are oxidized to CO and $CO_2$. As the $SiO_2$ layer forms, it impedes further oxidation of residual carbon atoms, leaving them at the $SiO_2$/SiC interface as interstitial carbon ($C_i$) and carbon clusters[35]. During the thermal oxidation, the $C_i$ can reduce near-surface $V_c$ through vacancy exchange mechanisms driven by concentration gradients, resulting in an increase in minority carrier lifetime. However, addressing deeper $V_c$ necessitates extending oxidation time and greater concentration gradients for effective vacancy exchange diffusion. The process is inherently challenging.

It is highly complex for hydrogen atoms to interact with silicon carbide and this interaction involves multiple mechanisms. As illustrated in Figure 6(d), During the initial stage of hydrogen annealing, hydrogen atoms can effectively passivate nearly all Vc through interstitial diffusion in a short time due to their small size. This explains why both our work and that of Kimoto's group reported a temporary increase in minority carrier lifetime, but with the presence of thermal instability[21, 36]. However, after prolonged hydrogen annealing, the minority carrier lifetime continues to increase, and thermal stability also improves. This phenomenon suggests the presence of an additional mechanism. As illustrated in Figure 6(e), this stage may be attributed to two factors that contribute to the significant reduction of mid-range carbon vacancies during hydrogen annealing:

(I) Prolonged low-pressure hydrogen annealing causes selective etching and leads to the graphitization of the 4H-SiC epilayer surface, providing a substantial amount of carbon atoms and a concentration gradient. A high concentration gradient facilitates the diffusion of carbon atoms through a vacancy exchange mechanism, leading to the elimination of carbon vacancies [37].

(II) Both epitaxial growth and thermal oxidation lead to the formation of non-diffusible carbon clusters. However, hydrogen atoms can decompose these carbon clusters[38]. This facilitates the elimination of deeper carbon vacancies within the epitaxial layer without requiring long-distance diffusion of carbon atoms.

The two factors mentioned above are the primary reasons for the reduction of carbon vacancies in the 4H-SiC epilayer after hydrogen annealing. Although carbon vacancies significantly decrease after hydrogen annealing, they are not completely eliminated. As shown in Figure 6(f), hydrogen annealing results in a smooth surface with a low concentration of carbon vacancies in 4H-SiC, while the remaining carbon vacancies will continue to be passivated by hydrogen atoms.

## Conclusions

The interactions between thermal oxidation, hydrogen annealing, and defects in 4H-SiC epilayers have been extensively studied. While thermal oxidation enhances minority carrier lifetime, it may also introduce interstitial carbon atoms and dislocation-related pits. High-temperature hydrogen annealing reduces stacking fault and dislocation density but may also lead to etch pit formation. In contrast, low-pressure low-temperature hydrogen annealing protects the 4H-SiC epilayer surface, significantly eliminates carbon vacancies, and improves minority carrier lifetime. The synergistic effects of these treatments create a dual mechanism that substantially enhances minority carrier lifetime. P-type 4H-SiC epilayers with exceptionally long minority carrier lifetimes have been achieved through a combination of double thermal oxidation and extended low-pressure hydrogen annealing. These findings contribute to a fundamental understanding of minority carrier dynamics in 4H-SiC and have important implications for improving the performance of SiC-based devices.

**Data availability statement**

All data that support the findings of this study are included within the article (and any supplementary files).

**Acknowledgments**

This research was supported by the National Natural Science Foundation of China (Grant No. 62274137), Natural Science Foundation of Jiangxi Province of China for Distinguished Young Scholars (No. S2021QNZD2L0013), the Fundamental Research Funds for the Central Universities (Grant No. 20720230103, Grant No. 20720220026), National Key Research and Development Program of China (Grant No.2018YFB0905700), the State Key Laboratory of Advanced Power Transmission Technology (Grant No. GEIRI-SKL-2022-005), Jiangxi Provincial Natural Science Foundation (Grant No. 20232BAB202043), the Science and Technology Project of Fujian Province of China (Grant No. 2020I0001).